\documentstyle[sprocl]{article}
\def\lt{Lense-Thirring}
\def\rp#1#2{{#1\over#2}}

\def\rfr#1{eq.(\ref{#1})}

\def\d{\delta}

\def\m{\mu}

\def\p{\pi}

\def\f{\phi}

\def\og{\omega} 
  
\def\D{\Delta}

\def\O{\Omega}  
\def\ct#1{\cite{#1}}
\def\lb#1{\label{#1}}
\def\bb{\bibitem}
\def\eqi{\begin{equation}}
\def\eqf{\end{equation}}
\def\gr{general relativity}
\def\lg{LAGEOS}
\def\lgg{LAGEOS II}
\def\grc{gravitomagnetic}

\def\pg{perigee}

\def\cl{clock}
\def\eft{effect}
\begin{document}

\title{SOLID AND OCEAN EARTH TIDES AND
THE DETECTION OF SOME GRAVITOMAGNETIC EFFECTS}
\author{ LORENZO IORIO}
\address{Dipartimento Interateneo di Fisica dell' Universit{\`{a}} di Bari,\\
Via Amendola 173, 70126, Bari, Italy\\E-mail: iorio@ba.infn.it.}

\maketitle\abstracts{The detection of some tiny  gravitomagnetic effects in the
field of the Earth by means of artificial satellites
is a very demanding task because of the various
other perturbing forces of gravitational and
non-gravitational origin
acting upon them. Among the gravitational perturbations  a relevant role is
played by the
Earth solid and ocean tides. In this communication I outline their effects on
the detection of the
Lense-Thirring drag of the orbits
of LAGEOS and LAGEOS II, currently analyzed, and  the proposed GP-C 
experiment devoted to the measurement of the clock effect.}

\section{The \lt\ drag of the orbits of the \lg s}
The \lt\
\eft~\ct{leti}~\ct{ciuwhe}
is currently measured~\ct{ciu2000} by
analyzing the following 
combination
of the orbital residuals of the nodes $\O$ of  \lg\ and
\lgg\ and the \pg\ $\og$ of \lgg~\ct{ciu96}:
\eqi
\d\dot\O^{I}+c_1\d\dot\O^{II}+c_2\d\dot\og^{II}\simeq
60.2\m_{LT},\lb{ciufor}\eqf
in which $c_1=0.295$, $c_2=-0.35$ and $\m_{LT}$ is the parameter
to be measured which is 1 in \gr\ and 0 in
Newtonian mechanics.
The \gr\ forecasts for the combined residuals a linear trend  with
slope of $60.2$ mas/y.

\subsection{The solid tides}
Concerning the solid tides, the most powerful
constituents turn out to be the semisecular 18.6-year, the $K_1$ and the
$S_2$ which induce
on the considered \lg s' elements  perturbations
of the order of $10^{2}-10^{3}$ mas with periods ranging from
111.24 days for the $S_2$ on \lgg\  to 6798.38 days for the
18.6-year tide.

The latter, with its amplitudes of
-1079.38 mas, 1982.16 mas and -1375.58 mas for the nodes of  the \lg s and
the
\pg\ of
\lgg\ respectively, could be
particularly insidious for
the detection of the \lt\ \eft. Indeed, since the observational periods
adopted until now~\ct{ciu2000} range only from  3.1 to 4 years, it
could resemble to a superimposed linear trend which may alias the recovery
of $\m_{LT}$. However, its effect should vanish since it is a $l=2\ m=0$
tide, and \rfr{ciufor} should be not sensible to such
tides~\ct{ciu96}. This feature
will be quantitatively assessed later.

Also the $K_1$ plays a not negligible role: it induces on \lg' s
node a perturbation with period of 1043.67 days and amplitude of
1744.38
mas, while
the node and the \pg\ of \lgg\ are shifted by an amount of -398 mas and
1982.14 mas, respectively, over a period of -569.21 days.
\subsection{The ocean tides}
About the ocean tides, whose  knowledge is less accurate
than that of the solid tides,
the $l=3$ part of the tidal spectrum turns out to be very interesting for
the
\pg\ of \lgg.

Indeed, for this element the $K_1$ $l=3\ p=1,2\
q=-1,1$ terms
induce perturbations which are
of the same order of magnitude of those generated by the solid tides, both
in the
amplitudes and in the periods.
E.g.,
the $l=3\ p=1\ q=-1$ harmonic has a period of 1851.9 days (5.09
years) and an
amplitude of 1136 mas, while the $l=3\ p=2\ q=1$ harmonic is less
powerful with its 346.6 mas and a period of -336.28 days.

It
should
be considered that, contrary to the
first two even degree zonal perturbations
which do not affect \rfr{ciufor}, the diurnal odd degree tidal
perturbations are not
canceled out by the combined residuals
of Ciufolini. So, over an observational period of few years the $K_1\ l=3\
p=1\ q=-1$
harmonic
may alias the \grc\ trend of interest.

The even degree ocean tidal
perturbations are not particularly
relevant: they amount to some tens of mas or less, with the
exception of
$K_1$ which perturbs the node of \lg\ and the \pg\ of \lgg\ at a
level of $10^2$ mas.
\section{The effect of the orbital tidal perturbations on the
measurement of the \lt\ \eft}
Given an observational period $T_{obs}$, the tidal perturbations must be
divided into two main categories according to their periods $P$: those
with
$P<T_{obs}$ and those with $P>T_{obs}$. While the former ones, even if
their
mismodeled amplitude is great so that they heavily affect the orbital
residuals, are not particularly insidious because their effect averages
out
if $T_{obs}=nP,\ n=1,2...$, the latter ones, on the contrary, are
particularly hazardous since they may alter the determination of $\m_{LT}$
acting as superimposed bias. This is particularly true for those diurnal
and semidiurnal tides which should affect the combined residuals, as the
$K_1\ l=3\ m=1\ p=1\ q=-1$. 

A preliminary analysis has been conducted by calculating \rfr{ciufor}
with the nominal tidal perturbative amplitudes worked out in the previous
sections for $T_{obs}$ =1 year. The calculations have been repeated also
with the mismodeled amplitudes. They show that, not only the $l=2,4\ m=0$
tides tend to cancel out, but also that this feature extends to the $l=3\
m=0$ ocean tides.

A more refined procedure will be described below for the diurnal and
semidiurnal tides.
\subsection{Case a: $P<T_{obs}$}
The effect of such class of perturbations has been evaluated as follows.

The orbital residuals curve has been simulated with MATLAB by including
the \lt\ trend as predicted by \gr, the main mismodeled tidal
perturbations
in the form of: \eqi \d A_{f}\sin{(\rp{2\p}{P_{f}}t+\f_{f})},\eqf where
$f$ denotes
the harmonic chosen,
and a
noise.
About the mismodeling $\d A$, we have assumed that the main source of
uncertainties are the free space potential Love number $k_2$~\ct{love} for
the solid
tides,
the load Love numbers~\ct{pag} and the ocean tidal
heights
$C^{+}_{lmf}$~\ct{pav} for the ocean tides.
The MATLAB routine has been endowed with the possibility of changing the
time
series length $T_{obs}$, the time sampling $\D t$, the amplitude of the
noise,
and the
harmonics' initial phases $\f$. 

The so
obtained simulated curves, for different choices of $T_{obs}$, have been   
subsequently fitted with a least-square
routine in order to recover, among the other things, the parameter
$\m_{LT}$.
This procedure has been repeated with and without the whole set of
mismodeled
tidal signals so to obtain an estimate
$\D\m_{tides}=\m_{LT}(all\ tides)-\m_{LT}(no\ tides)$ of their influence
on
the measurement of $\m_{LT}$. This analysis show that
$2\%<\D\m_{tides}<4\%$
for
$T_{obs}$ ranging from 4 years to 7 years.

\subsection{Case b: $P>T_{obs}$}
The effect of such class of perturbations has been evaluated with
different
approaches.

Firstly, in a very conservative way, it has been considered the averaged
value of
the mismodeled tidal signal under consideration over different $T_{obs}$.
The analysis has been performed for the 18.6-year tide and the $K_1\ l=3\  
p=1\ q=-1$.
On the
semisecular
tide it has been assumed a mismodeling level of $1.5\%$
due to the uncertainty at its frequency of the anelastic behavior of
the Earth' s mantle accounted for by the Love number $k_2$.
The oceanic constituent has been considered unknown at a level of almost
$6\%$ due to the uncertainties on the load Love number of degree $l=3$
and to the tidal height coefficient $C^{+}_{lmf}$ as released by EGM96.
Over different $T_{obs}$ the latter  
affects the determination of the \lt\ effect at a level of $2.3\%$ at
most,
while the
zonal tide, as predicted by Ciufolini~\ct{ciu96}, almost cancels out
giving
rise
to negligible contributions to the combined residuals.

\section{The clock effect}
In the \grc\ \cl\ \eft~\ct{mash2}~\ct{mash3}~\ct{tart1}~\ct{tart2}
two clocks moving along pro- and
retrograde circular equatorial
orbits, respectively, about the Earth exhibit a difference in
their proper times which, if calculated after some fixed angular
interval, say $2\p$, amounts
to almost $10^{-7}$ s

In~\ct{mash3}~\ct{gron}
it has been shown that
for an orbit radius of 7000 km the radial
and azimuthal locations of the satellites must be known at a level of
accuracy of $\d r\leq
10^{-1}$ mm and $\d \f\leq 10^{-2}$ mas per revolution.

\subsection{The systematic radial errors induced by the tidal perturbations}
About the radial direction, the even degree part of the tidal spectrum does not affect it
contrary to the odd degree part~\ct{iorio}. For the $l=3\ m=1\ p=1\ q=-1$
ocean tides we have:
\eqi\D r_{311-1f}=(8.80\cdot 10^{25}\ cm^{7/2}\
s^{-1})\times a^{-7/2}\times P_{pert}\times C^{+}_{31f},\lb{radoc}\eqf
where $a$ is the satellite' s semimajor axis and  $P_{pert}$ is the
perturbation' s period.

For $K_1$, the most powerful diurnal ocean tide, we have $P_{pert}=50$ days and $\d
r_{311-1}(K_1)\simeq 2.07$ cm by assuming an uncertainty of $5.2\ \%$ on the tidal height
coefficient~\ct{pav}. Despite 
the amplitude of this long period mismodeled
perturbation is 2 orders of
magnitude
greater than the maximum allowable error $\d r_{max}= 10^{-1}$ mm,
it must be noted that its
period $P_{pert}$ amounts to only 50 days. This implies that if an observational time span $T_{obs}$   
which is an integer multiple of $P_{pert}$, i. e. some months, is adopted the tidal perturbative
action
of $K_1$ can be averaged out.

\subsection{The systematic azimuthal errors induced by the tidal perturbations}

The azimuthal angle is perturbed only by the even part of the tidal
spectrum~\ct{iorio}. For $l=2$ we have:
\eqi \D\f^{solid}=(-3.77\cdot 10^{18}\ cm^{5/2}s^{-1}) \times
a^{-7/2}\times
k^{(0)}_{20}\times P_{pert}\times H^{0}_{2},\lb{cazs}\eqf
\eqi \D\f^{ocean}=(-4.707\cdot 10^{17}\ cm^{5/2}s^{-1}) \times
a^{-7/2}\times   
P_{pert}\times C^{+}_{lmf},\lb{cazzarol}\eqf where
$H_{l}^{m}$ are the Doodson coefficients with a different
normalization~\ct{mcca}~\ct{roo}.
For $a=7000$ km we have:\\
$\bullet\ \D\f(18.6-year)=-4.431\cdot 10^{4}$ mas\\
$\bullet\ \D\f(9.3-year)=-214.4$ mas\\
$\bullet\ \D\f(S_a)=408$ mas (solid); 857.6 mas (oceanic)

The zonal tidal perturbations on the satellite' s azimuthal location are
particularly insidious not only because their nominal amplitudes are up to 6
orders of magnitude greater than the maximum allowable error
$\d\f_{max}=10^{-2}$ mas, but also
because they have periods very long, so that there is no hope they average out
on reasonable $T_{obs}$. Concerning the 18.6-year tide, by
assuming an uncertainty
of $1.5\%$ on $k^{(0)}_{20}$, the mismodeling
on its perturbation amounts to
-664 mas which is, however, very far from $\d\f_{max}$.

\section*{Acknowledgments}
I am profoundly indebted to Prof. I. Ciufolini, who directed my attention to the topic of
the orbital perturbations, to Prof. E. Pavlis for his patience and helpful discussions and to
Prof. L. Guerriero who supported me at Bari. I thank also Prof. B. Mashhoon for the useful material
supplied to me. I would also like to thank the organizers of the Spanish Relativity Meeting (EREs2000) for
their
warm hospitality.

\end{document}